\documentclass[conference]{IEEEtran}
\IEEEoverridecommandlockouts
\usepackage{cite}
\usepackage{amsmath,amssymb,amsfonts}
\usepackage{algorithmic}
\usepackage{graphicx}
\usepackage{textcomp}
\usepackage{xcolor}
\usepackage{graphicx}
\usepackage{subcaption}
\usepackage{tabularx}
\usepackage{booktabs}
\usepackage[utf8]{inputenc}
\usepackage{hyperref}
\def\BibTeX{{\rm B\kern-.05em{\sc i\kern-.025em b}\kern-.08em
    T\kern-.1667em\lower.7ex\hbox{E}\kern-.125emX}}
\begin{document}

%\title{A Machine Learning Model to Detect Alcohol Influenced Driving Using Mobile Phone Sensors’ Data*\\
\title{Mobile Phone Sensor-based Nigerian Driving Dataset to Detect Alcohol-influenced Behaviours\\
}

\author{
\IEEEauthorblockN{Iniakpokeikiye Peter Thompson}
\IEEEauthorblockA{\textit{Department of Computing Science} \\
\textit{University of Aberdeen}\\
Aberdeen, United Kingdom \\
i.thompson.21@abdn.ac.uk}
\and
\IEEEauthorblockN{Yi Dewei}
\IEEEauthorblockA{\textit{Department of Computing Science} \\
\textit{University of Aberdeen}\\
Aberdeen, United Kingdom \\
dewei.yi@abdn.ac.uk}
\and
\IEEEauthorblockN{Reiter Ehud}
\IEEEauthorblockA{\textit{Department of Computing Science} \\
\textit{University of Aberdeen}\\
Aberdeen, United Kingdom \\
e.reiter@abdn.ac.uk}

}

\maketitle

\begin{abstract}
This paper presents a unique driving dataset collected in Nigeria via mobile phone sensors to support a machine learning model for detecting alcohol-influenced driving behaviours, with the long-term aim of integrating this model into a mobile application that encourages safer driving behaviours. We adopt existing data processing and pattern-matching methodologies to label real-world driving data collected from Nigerian drivers, and are used to train the model. A decision tree classifier is developed to detect alcohol influence, based on behavioural and temporal features, achieving a recall of 100\%, a precision of 60\%, and an F1 score of 75\%. The model’s overall accuracy was 90.91\%, ensuring that no alcohol-influenced trips were missed. Key predictive features include speed variability, course deviation, and time of day, which align with established patterns of alcohol consumption. This study contributes to the field by demonstrating how machine learning can be applied in low-resource environments to improve road safety. The findings suggest that the model can significantly enhance the detection and prevention of risky driving behaviours, with the potential for future integration into mobile applications to provide real-time feedback that encourages safer driving practices. This scalable and accessible solution offers a new approach to addressing road safety challenges in regions where traditional interventions are inadequate.
\end{abstract}


\section{Introduction}
Driving under the influence of alcohol continues to pose a significant threat to public safety worldwide, leading to impaired coordination of actions including reduced judgment, and delayed reaction times. These impairments dramatically increase the likelihood of traffic accidents and fatalities. Although traditional methods for detecting alcohol-influenced driving, such as breathalysers and roadside sobriety tests are effective, they are limited by the requirement of law enforcement presence and driver compliance. This limitation underscores the need for innovative, noninvasive, and continuous monitoring solutions capable of detecting impaired driving behaviours in real-time, particularly in contexts where law enforcement is less accessible.
The increasing ubiquity of smartphones equipped with advanced sensors, including accelerometers, gyroscopes, GPS, and others, provide a promising avenue for analysis and identification of drivers driving patterns and behaviours, especially with those associated with alcohol impairment, allowing for the detection of alcohol-influenced driving, without relying on external devices or law enforcement presence, when Machine Learning (ML) tools are employed.
This paper introduces a machine learning model designed to detect alcohol-influenced driving behaviour using data collected from smartphone sensors. The contributions of this study are outlined as follows:
\begin{itemize}
\item Lablling of naturalistic drivine dataset: This study tailors established methodologies to label and analyse real-world driving data collected from Nigerian drivers, addressing the unique geographical and behavioural context of the region.

\item Development of a machine learning model for detecting alcohol-influenced driving: The model uses sensor data from smartphones to accurately identify driving behaviours indicative of alcohol impairment, focusing on patterns unique to Nigerian drivers.

\item Bridging the gap between theoretical model development and practical application: By focusing on real-world driving data and developing a model that can be integrated into mobile technologies, this study moves beyond theoretical frameworks to offer a practical solution for enhancing road safety.
\end{itemize}

\section{Related Work}

\subsection{Mobile Phone Sensors in Behavioural Analysis}
Mobile phone sensors have emerged as powerful tools in behavioural analysis, offering a nonintrusive and efficient means to gather extensive data on human activities and traits. Researchers have been able to collect a comprehensive dataset from participants' phones, including accelerometer readings, GPS locations, and other movement data. 
In driving behaviour analysis, research has focused on the use of mobile phone sensors for the collection of data used to analyse drivers’ driving behaviours. In practice, a comprehensive dataset using accelerometer and gyroscope readings has been collected from multiple drivers over seven days using mobile phone sensors such as accelerometer and gyroscope \cite{wawage2022datasetdriver}. This data was captured using the "Sensor Record" Android application, which recorded various driving parameters including speed, acceleration, braking, steering, and location data under realistic traffic conditions. The dataset is aimed to facilitate the training, testing, and validation of machine learning models for driver behaviour classification into normal, aggressive, and risky driving patterns. A significant limitation in this study is the variability in data quality due to different smartphone models and sensor sensitivities because the accuracy of the sensors can vary significantly between devices, leading to inconsistencies in data collection.

\subsection{Alcohol Use and Alcohol Influenced Driving in Nigeria}
Alcohol-impaired driving has been consistently identified as a significant contributor to unsafe driving behaviours and road traffic accidents. According to \cite{pradeep2020roadaccidents}, alcohol intake is a major human factor responsible for road traffic accidents. In Nigeria, \cite{awosusi2021alcohol} found a strong negative correlation between alcohol use and compliance with road safety rules among commercial motorcyclists. This is further underscored in \cite{uhegbu2021roadsafety} study, which reported that about one-third of road users in Abuja, Nigeria, admitted to driving under the influence of alcohol. Additionally, research on the impact of psychoactive substance use on road rage behaviour among commercial drivers in Nigeria identified substances like alcohol, cocaine, and amphetamine as predictors of road rage behaviour \cite{uzondu2019roadsafety}. 

Driving Under the Influence (DUI) is a significant issue in Nigeria, with studies highlighting various aspects related to substance abuse among drivers \cite{nwakobi2022forensic, osamika2021substanceuse}. Commercial drivers, especially, are prone to using psychoactive substances to cope with long working hours, leading to increased risk of accidents \cite{akande2023psychoactive, osamika2021substanceuse}. Nigeria does not have a law passed by its parliament with regards to legal Blood Alcohol Concentration (BAC) level. Notwithstanding, The Federal Road Safety Corps (FRSC) of Nigeria set a BAC limit of 0.05g/100mL, but its enforcement is weak due to the unavailability of BAC testing equipment. This further underscores why there is advent scarcity of data on drivers, their BAC concentration levels and corresponding driving behaviour data. Additionally, the lack of legislation on (BAC) in Nigeria contributes to the prevalence of drunk driving incidents \cite{akande2023psychoactive}. Driving Under Influence (DUI) is a root cause of other unsafe driving behaviours because an impaired driver is prone to commit many other offenses. 
To combat DUI effectively, interventions targeting drivers' socio-demographic factors, knowledge, and perception of substance abuse are recommended. Implementing and enforcing laws against drunk driving, along with leveraging forensic science in criminal investigations are part of the recommendation to reduce road accidents and fatalities related to DUI in Nigeria. If an alcohol influenced driving detector is implemented, it will enable an easy enforcement of DUI laws thereby reducing road rage and other unsafe driving behaviours, it is in this context that this study becomes necessary and it is the motivation for this work because this study pioneers a cheaper, more accessible, and non-intrusive way of detecting alcohol influenced driving in Nigeria at this time where there is no known cheap and easily accessible intervention.
\section{Methodology}
This study employs data mining, using statistical and machine learning techniques to analyze patterns in alcohol-influenced driving from real-world reports. The goal is to train and test baseline models on this data. While methods vary across studies depending on data characteristics, outcomes also differ by dataset and location. This study uses previously collected local data on reported influences on driven trips from Nigerian drivers \cite{peter2024telematicapp}. 
In this study, this data is pre-processed, features extracted, model trained and evaluated. This is even so that, researchers can use real-world, non-evasive driving data from countries like Nigeria to build and extend studies on alcohol-influenced driving.

\begin{figure*}[htbp]
  \centering
  \begin{subfigure}[t]{0.42\textwidth}
    \centering
    \includegraphics[width=1.2\linewidth]{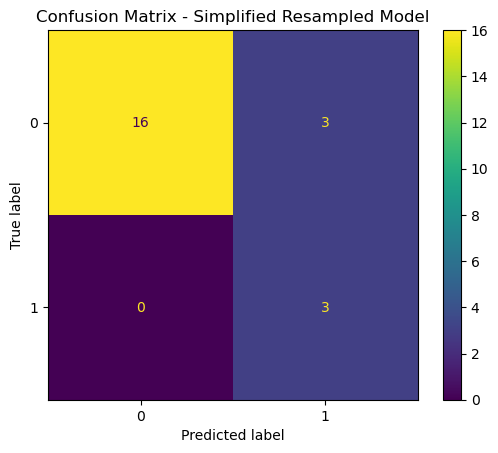}
    \caption{Confusion Matrix for Alcohol-Influenced Behaviours}
    \label{fig:confusion-matrix}
  \end{subfigure}
  \hspace{0.1\textwidth}
  \begin{subfigure}[t]{0.42\textwidth}
    \centering
    \includegraphics[width=1.1\linewidth,height=3.3in]{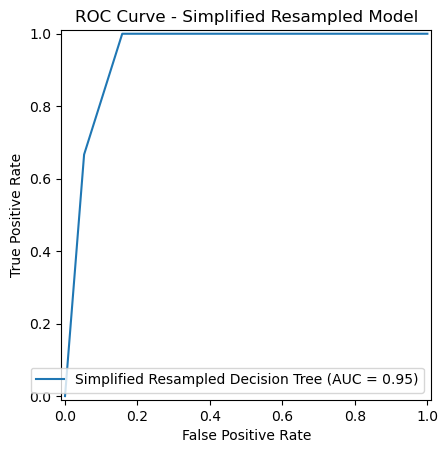}
    \caption{ROC Curve for Model’s Class Discrimination Ability}
    \label{fig:roc-curve}
  \end{subfigure}
  \caption{Model Evaluation Metrics: (a) Confusion matrix for alcohol-influenced behaviours; (b) ROC curve for model’s class discrimination ability.}
  \label{fig:model-evaluation}
\end{figure*}

\subsection{Data Collection}
A Telematic Data Collection App for Android phones; powered by Damoov telematic SDK was used for the data collection (Peter et al., 2024). This app was installed on 5 Techno Pop 6 with Android 11, Quad Core CPU mobile phones and was given to drivers on a rotational basis. The phones were vetically mounted on each driver’s vehicle screens to collect real-time driving data. Data was sent to the Damoov datahub for daily download.
The app starts data collection when the driver taps on the 'Start Trip' button on the app. Data collection is stopped when driver taps on the 'End Trip' button. A survey form is activated by tapping ‘End Trip’ after the end of a trip. The Data were collected from 21 drivers, including 5 private and 16 public drivers were collected over a three-month period.
\subsection{Dataset}
The dataset comprises 436,715 sensor data points from 230 trips, including gyroscope, accelerometer (x, y, z), GPS (latitude, longitude), speed, distance, time, and driver-reported trip influences.

\begin{itemize}
\item{\textit{Understanding the Data}}\\
Data includes mobile sensors' readings and self-reported influences via an in-app survey triggered at trip end. Drivers reported alcohol intake, overloading, or other influences, along with a timestamp. While various methods like olfactory sensors, thermal imaging, and biosensors \cite{li2024drunksafetyreview, bhawarthi2023alcoholdetector} have been used in literature, they are often intrusive, expensive, and legally constrained in countries like Nigeria.

\item{\textit{Sensor Data}}\\
The app collected real-time accelerometer, gyroscope, GPS, distance, time, and driver ID data at 1 Hz. Prior studies show accelerometer and GPS data are effective in detecting alcohol-influenced driving \cite{chawathe2020bacestimation, wu2023sensorcraving, brahim2022driverclassification}. These commonly available smartphone sensors offer a practical, low-cost, and non-intrusive alternative.
\end{itemize}

\subsection{Data Pre-Processing}
The raw dataset include 463,384 data points across 230 trips, with features such as driver type, distance, speed, location (latitude, longitude), timestamp, acceleration, and self-reported trip influences (e.g., alcohol, fatigue, phone use).

Data cleaning involved removing trips without driver reports or with invalid characters in the influence column. Features like lateral and yaw were dropped due to excessive missing values. The cleaned dataset contains 207,741 data points from 108 trips.

The \texttt{influence} column was binary-encoded: alcohol = 1, all other influences = 0. Driver type was encoded as public = 1 and private = 0, ensuring all features were numeric for analysis.

\begin{table}[htbp]
\caption{Dataset Feature Descriptions}
\centering
\renewcommand{\arraystretch}{1.1}
\begin{tabular}{|p{1.4cm}|p{3.2cm}|p{3.0cm}|}
\hline
\textbf{Feature} & \textbf{Description} & \textbf{Sensor Values} \\
\hline
Driver & Driver ID and type. & ID, type \\
\hline
Distance & Distance travelled (meters). & \texttt{totalMeters} \\
\hline
GPS & Location coordinates. & Latitude, longitude \\
\hline
Speed & Speed readings. & \texttt{speed}, \texttt{midSpeed} \\
\hline
Accel. & Motion along X, Y, Z axes. & \texttt{accX}, \texttt{accY}, \texttt{accZ} \\
\hline
Date & Date and time (string). & Date, time \\
\hline
Direction & Heading from North. & \texttt{course} \\
\hline
Altitude & Vehicle altitude. & \texttt{height} \\
\hline
Timestamp & Data point time. & \texttt{tickTimestamp} \\
\hline
Track ID & Trip identifier. & \texttt{tripId} \\
\hline
Influence & Reported trip factors. & E.g., traffic, weather \\
\hline
\end{tabular}
\label{tab1}
\end{table}

\subsubsection{Feature Extraction/Selection}
Following preprocessing, the \texttt{point\_date} feature was decomposed into \texttt{time\_of\_day} and \texttt{day\_of\_week} to align with typical drivers alcohol consumption patterns observed in Nigeria.

Given the trip-based data structure, alcohol influence detection was modeled at the trip level not the point level. As in similar studies \cite{bhowmik2019aggregation, pellegrini2021aggregation, zhao2024timefrequency}, trip-level data was aggregated using statistical summaries (mean, min, max, standard deviation) to capture general driving behavior while reducing noise.

To refine the aggregated dataset, multiple feature selection techniques were applied to identify the most relevant features. These included statistical and machine learning-based methods: Select K-Best (SKB), Recursive Feature Elimination with Cross-Validation (RFECV), Select Percentile, Principal Component Analysis (PCA), and Random Forest Feature Importance.

Each method selected key features based on different criteria such as statistical relevance or predictive strength. Table~\ref{tab:feature_selection} shows the features selected by each technique.

\begin{table}[htbp]
\caption{Feature Selection (FS) Methods and Selected Features}
\begin{center}
\begin{tabular}{|p{3.0cm}|p{4.1cm}|}
\hline
\textbf{FS Method} & \textbf{Selected Features} \\
\hline
Select Percentile & Driver type, longitude mean, longitude min, longitude max, longitude std, course std, tickTimestamp mean, tickTimestamp max, tickTimestamp min, accelerationYOriginal min, day of week mean \\
\hline
Principal Component Analysis (PCA) & totalMeters min, totalMeters mean, speed mean, driver type, speed min, midSpeed std, speed max, midSpeed min, speed std, totalMeters max \\
\hline
Select K-Best (SKB) & Driver type, longitude mean, longitude min, longitude max, longitude std, height mean, tickTimestamp mean, tickTimestamp max, tickTimestamp min, day of week mean \\
\hline
Random Forest Feature Importance & accelerationYOriginal mean, longitude min, longitude mean, longitude max, tickTimestamp max, tickTimestamp min, height mean, course std, tickTimestamp mean, height max \\
\hline
Recursive Feature Elimination with Cross-Validation & Latitude min, longitude min, tickTimestamp min, accelerationYOriginal mean \\
\hline
\end{tabular}
\label{tab:feature_selection}
\end{center}
\end{table}

While algorithmic feature selection is valuable, incorporating domain knowledge enhances model relevance, especially for context-specific deployments like this Nigerian case study. Conversations with local drivers revealed that alcohol consumption is more common at certain times and days, guiding the inclusion of temporal features.

Literature also identifies key driving behaviors affected by alcohol:

\begin{itemize}
    \item \textit{Lane Positioning Impairment:} Drivers with BAC levels of 0.10 struggle to maintain lane position \cite{nhtsa2022drunkdriving}.
    \item \textit{Speed Control Variability:} At a BAC of 0.08, consistent speed becomes difficult to maintain \cite{nhtsa2022drunkdriving}.
\end{itemize}
These BAC levels exceed Nigeria's legal limit of 0.05 \cite{chen2016bacsimulator, kumar2022reviewiotml}.
Based on this, the following feature categories were emphasized:

\begin{itemize}
    \item \textit{Temporal Features:} Mean hour and day of the week capture patterns associated with peak alcohol use periods in Nigeria.
    \item \textit{Behavioral Features:} 
    \begin{itemize}
        \item Standard deviation of speed reflects impaired speed control.
        \item Standard deviation of course captures lane maintenance difficulty.
    \end{itemize}
    \item \textit{Physiological Features:} 
    \begin{itemize}
        \item Mean lateral acceleration (Y-axis) indicates acceleration and braking irregularities tied to delayed reaction times.
    \end{itemize}
\end{itemize}
These domain-informed features are summarized in Table~\ref{tab:domain_features}.

\begin{table}[htbp]
\caption{Description of Key Selected Features}
\begin{center}
\begin{tabular}{|p{3.0cm}|p{4.1cm}|}
\hline
\textbf{Feature} & \textbf{Description} \\
\hline
\textbf{speed\_mean} & Mean speed during trips. \\
\hline
\textbf{acceleration\_std} & Standard deviation of overall acceleration, indicating variability. \\
\hline
\textbf{course\_std} & Variability in vehicle heading, reflecting directional changes. \\
\hline
\textbf{hour\_of\_day\_mean} & Mean hour of the day when trips occur, correlating with potential alcohol consumption times. \\
\hline
\end{tabular}
\label{tab:domain_features}
\end{center}
\end{table}

\subsubsection{Model Design}
Following data preprocessing, the dataset reduced to 108 trips as stated earlier with 14 alcohol-influenced and 94 non-alcohol-influenced data, resulting in significant class imbalance. A Decision Tree Classifier was trained, and the Synthetic Minority Oversampling Technique (SMOTE) was applied on the dataset to address the issue of data imbalance.The data set was divided into 80\% training set and a 20\% testing set. The primary objective was to maximize recall and maintain a strong F1-score, ensuring that no alcohol-influenced cases were missed.

\begin{table}[htbp]
\caption{Performance Comparison of Classification Models Using Resampled Data}
\centering
\begin{tabularx}{\columnwidth}{@{}l>{\centering\arraybackslash}X>{\centering\arraybackslash}X>{\centering\arraybackslash}X>{\centering\arraybackslash}X>{\centering\arraybackslash}X@{}}
\toprule
\textbf{Model} & \textbf{Acc.} & \textbf{Prec.} & \textbf{Recall} & \textbf{F1} & \textbf{AUC} \\
\midrule
Logistic Regression & 0.852 & 0.70 & 0.75 & 0.72 & 0.80 \\
SVM (RBF Kernel) & 0.881 & 0.78 & 0.80 & 0.79 & 0.85 \\
Random Forest & 0.905 & \textbf{0.85} & 0.92 & \textbf{0.88} & 0.93 \\
\textbf{Decision Tree} & \textbf{0.909} & 0.60 & \textbf{1.00} & 0.75 & \textbf{0.95} \\
\bottomrule
\end{tabularx}
\label{tab:model_comparison}
\end{table}

\subsection{Model Selection and Training}
We needed to carefully select a model that could be the best representative model for the study case. We trained and evaluated a range of machine learning models, including Logistic Regression, Support Vector Machines (SVM), Random Forest, and a Decision Tree classifier.

\subsubsection{Evaluation Metrics}
The performance of the models was evaluated on a test data set using various metrics, including accuracy, precision, F1 score, and Receiver Operating Characteristic (ROC). Although the Random Forest model achieved slightly higher accuracy as presented in Table \ref{tab:model_comparison} , we ultimately selected the Decision Tree 
\begin{figure*}[htbp]
\centering
\includegraphics[width=19.1cm, height=8.7cm]{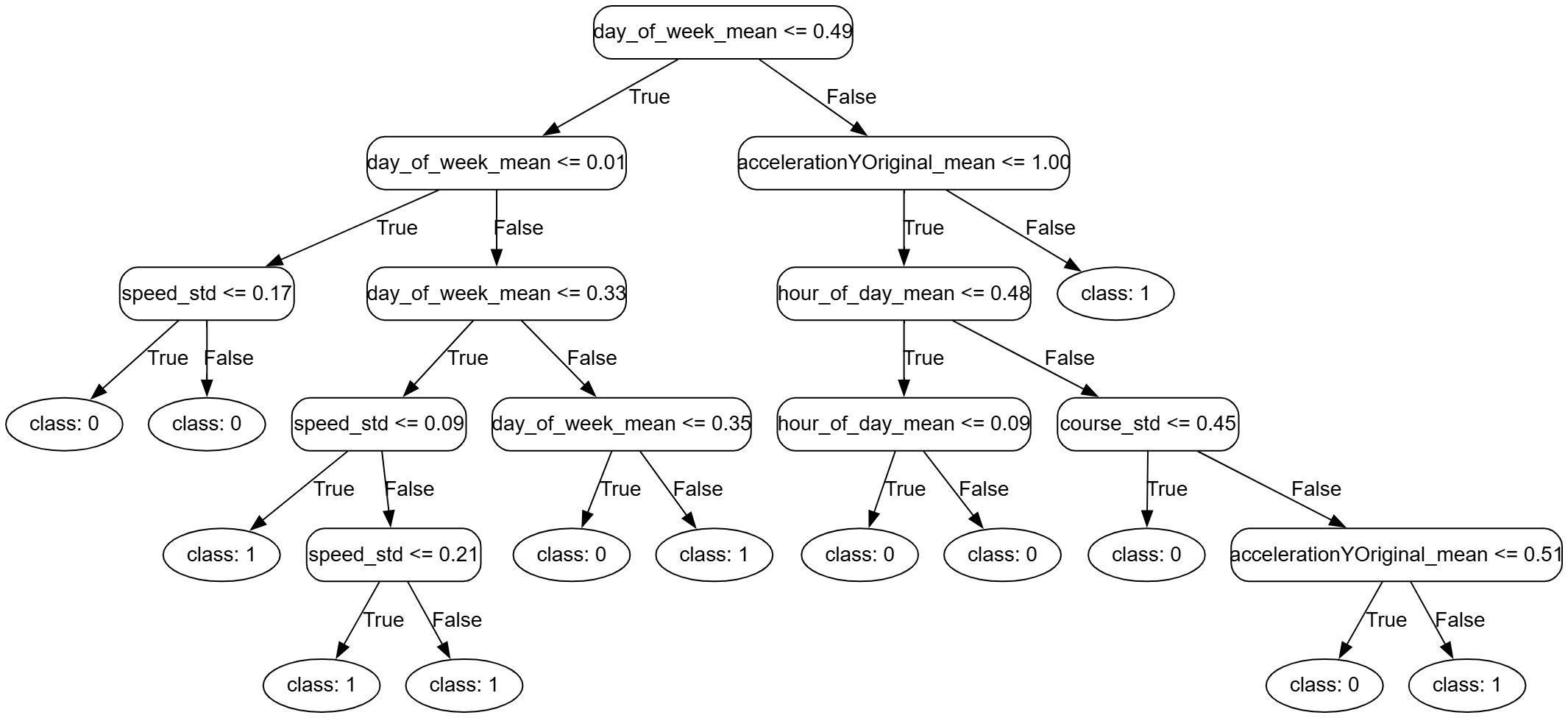}
\caption{Decision Tree Visualization of the Alcohol-Influenced Driving Behaviour Detection Model}
\label{fig:dt}
\end{figure*}
classifier for two main reasons:
\begin{enumerate}
    \item \textit{High Recall:} The Decision Tree model achieved a recall of 100\% for the alcohol-influenced class, which means it correctly identified all instances of drunk driving. In a safety-critical application like this, it is far more important to avoid false negatives (missing a drunk driver) than to have a few false positives.
    \item \textit{Interpretability:} Decision Trees are highly interpretable, which is a significant advantage in this context. The ability to understand why the model is making a particular prediction can help to build trust in the system and can also provide insights into the key indicators of drunk driving.
\end{enumerate}
These results demonstrate that the model effectively prioritized recall to ensure the detection of all alcohol-influenced driving cases, even at the expense of moderate precision.

The Decision Tree model was pruned and optimized to prevent overfitting.  The hyperparameters were set as follows: \texttt{max\_depth=5}, \texttt{min\_samples\_split=10}, and \texttt{min\_samples\_leaf=5}.  A \texttt{random\_state=42} was used to ensure reproducibility.

\subsubsection{Decision Tree Plot}

The decision tree starts with day of week mean as the root feature, reflecting the insight that alcohol consumption among Nigerian drivers occurs more frequently on specific days, such as weekends. This is followed by hour of day mean at many nodes, capturing the higher likelihood of alcohol influence during certain times, such as late nights. As the tree branches further, behavioural features like speed standard deviation and course standard deviation emerge, indicating impaired speed control and difficulties in maintaining lane position, both are well-documented effects of alcohol consumption. Finally, accelerationY Original mean highlight’s reaction delays in swerving, reinforcing the model’s alignment with both domain knowledge and established research on alcohol-impaired driving behaviours. The decision tree structure mirrors the temporal patterns and driving behaviours critical for detecting alcohol influence. The decision tree is presented in Figure~\ref{fig:dt}.

\section{Results and Discussion}

The decision tree classifier effectively detected alcohol-influenced driving behaviour, with key predictors \texttt{day\_of\_week\_mean}, \texttt{hour\_of\_day\_mean}, \texttt{speed\_std}, and \texttt{course\_std}—closely aligning with known alcohol-related driving patterns among Nigerian drivers, as visualized in Figure~\ref{fig:dt}.

The confusion matrix (Figure~\ref{fig:confusion-matrix}) illustrates the model's performance on 22 test cases: 16 true negatives and 3 true positives, with 3 false positives and no false negatives. This means all alcohol-influenced trips were correctly detected; an essential factor for road safety.

The model achieved:
\begin{itemize}
    \item \textbf{Precision:} 60\% (3 out of 5 predicted alcohol-influenced trips were correct)
    \item \textbf{F1 Score:} 75\%, balancing precision and recall
    \item \textbf{Accuracy:} 90.91\%, indicating strong overall performance
\end{itemize}

The ROC curve (Figure~\ref{fig:roc-curve}) reports an AUC of 0.95, demonstrating high discriminative power in distinguishing alcohol-influenced from non-alcohol-influenced trips.

The model's strong recall is particularly valuable in safety-critical applications, where failing to detect an alcohol-influenced driver could have serious consequences. The use of temporal features (e.g., weekend nights) and behavioural indicators (e.g., variability in speed and course) reflects days and times in the week when drivers most likely engage in drink driving as well as corresponding well-documented effects of alcohol on driving respectively.

While three trips were falsely flagged as alcohol-influenced (false positives), this is an acceptable trade-off given the priority of recall. In real-world deployments, such caution can help prevent accidents even at the cost of some false alarms.

Overall, the combination of domain-informed feature selection and statistical techniques led to a robust and practical model for detecting alcohol-influenced driving in Nigerian traffic conditions.

\section{Conclusion and Future Work}

This study successfully developed a machine learning model capable of detecting alcohol-influenced driving behaviours using smartphone sensor data, with a focus on deployment in resource-limited environments like Nigeria. Leveraging temporal and behavioural features, the decision tree classifier achieved an accuracy of 90.91\%, recall of 100\%, and an F1 score of 75\%. These results highlight the model's ability to reliably detect alcohol-impaired driving without missing critical cases; an essential factor for enhancing road safety. To our knowledge, this is the first documented drink driving dataset in Nigeria based on our findings.

Key predictive features, such as speed variability, and course deviation, align with documented patterns of alcohol-related impairment. In addition, this study also shows that other temporal features such as day of week and time of day are also viable candidate features for a model of this type, especially one that is targeted at the Nigerian types of culture. This confirms the effectiveness of using domain-informed features tailored to local driving contexts. The study also demonstrates the potential of machine learning in addressing road safety challenges in low-resource settings.

For future development, several areas are identified below:

\begin{itemize}
    \item \textbf{Improving Precision:} While recall is high, precision (60\%) can be improved by incorporating features that capture fine-grained driving behaviours; e.g., momentary speed deviations, abrupt braking, or swerving.

    \item \textbf{Real-Time Feedback Integration:} The ultimate goal is to integrate this model into an AI/NLG-enabled mobile application. In this system, the model will classify each trip in real time and feed the results into a natural language feedback mechanism. This feedback will help drivers understand and correct risky behaviours, encouraging safer driving habits.

    \item \textbf{Behaviour Change Support:} This AI-driven feedback loop offers a scalable alternative to traditional interventions, particularly valuable in low-income settings like Nigeria, where road safety enforcement and education are often limited.

    \item \textbf{Dataset Expansion:} Broader data collection is recommended to improve model generalizability. Including more regions, diverse road conditions, vehicle types, and driving patterns will make the model more robust and adaptable to varied environments.

    \item \textbf{Public Safety and Education Programs:} Integration of the model into government or NGO-led road safety initiatives could amplify its impact, enabling targeted interventions and contributing to a data-driven approach to driver education and accident prevention.
\end{itemize}

This research lays a solid foundation for scalable, AI-powered road safety solutions in under-resourced environments, with strong potential for real-world impact.

\section*{Acknowledgment}

Special thanks to the Tertiary Education Trust Fund (TetFund) for funding the bigger PhD research project through which this research is born. Also, special thanks to Sanu Momodu Kabiru, who has been very special in helping out as a research assistant with the data collection process and other research tasks during this research.

\bibliographystyle{ieeetr}
\bibliography{biblio}
\end{document}